# Quantum sensing of Lanthandie binding tags with relaxometer of NV center in diamond

Zibo Gao[1,2,3]†, Zhengzhi Jiang[4]†, Qiyu Liang[1,5]†, Ruihua He[1,5], Van Cuong Mai[6], Yingwei Tang[6], Qirong Xiong[6], Wenting Zhao[6], Hongwei Duan[6], Hongliang Sun*[7], Mo Li*[2,3], Yansong Miao*[5,8], Weibo Gao*[1,9,10]

[1]*Division of Physics and Applied Physics, School of Physical and Mathematical Sciences, Nanyang Technological University; Singapore, 637371, Singapore*

[2]*The Key Laboratory of Bionic Engineering, Jilin University, Changchun, 130022, China*

[3]*The College of Biological and Agricultural Engineering, Jilin University, Changchun, 130022, China*

[4]*Department of Chemistry, National University of Singapore, Singapore, 117543, Singapore*

[5]*School of Biological Sciences, Nanyang Technological University, Singapore, 637551, Singapore*

[6]*School of Chemistry, Chemical Engineering and Biotechnology, Nanyang Technological University, Singapore, 637457, Singapore*

[7]*Yunnan International Joint Laboratory of Bionic Science and Engineering，Kunming 650223, China.*

[8]*Institute for Digital Molecular Analytics and Science, Nanyang Technological University, Singapore, 636921, Singapore*

[9]*School of Electrical and Electronic Engineering, Nanyang Technological University, Singapore, 639798, Singapore*

[10]*Centre for Quantum Technologies, National University of Singapore, Singapore, 117543, Singapore*

*Corresponding author. † These authors contributed equally to this work.

Email: yansongm@ntu.edu.sg, moli@jlu.edu.cn, hlsun@aliyun.com, wbgao@ntu.edu.sg

Lanthanide binding tags (LBTs) stand out as a prominent group of fluorescent probes that are extensively utilized in biological detection. However, research on LBTs has predominantly emphasized their fluorescence properties, which frequently compromised by background fluorescence noise. Investigating magnetic properties could optimize detection methodologies that offer enhanced sensitivity and specificity. In this study, we measured the response of a relaxometer based on ensemble nitrogen-vacancy (NV) centers in diamond to various amounts of LBTs with gadolinium ions, determining the detection limit of LBTs to be 25 fmol. We then proposed and demonstrated a detection scheme employing the NV relaxometer to detect specific binding between LBTs and target. Specifically, we assessed the relaxometer's response to various concentrations of the interaction between the modified LBTs and Receptor-Binding Domain (RBD) of SARS-COVID-2 spike protein, with the detection threshold reaching ~1 pmol. Our research provides a potential application platform for biomarker detection under picomole concentration by using NV centers to detect the magnetism of LBTs.

1. Introduction

Sensitive and accurate detection of biomolecules is the cornerstone of experimental sciences, playing a curial role in disease diagnosis, drug discovery, food chemistry, and environmental sciences. With advancements in science and technology, the methods for in vitro biomolecule sensing have also seen significant and rapid improvements[1]. For example, the single-molecule surface-induced fluorescence attenuation (smSIFA) method[2] is utilized to explore the interactions and molecular dynamics between biomolecules and lipid bilayers[3]. However, these optically based methods encounter challenges in improving sensitivity due to the background fluorescence[4]. Biomolecular detection based on magnetic properties provides an effective approach to overcome background issues in recent years. For instance, by leveraging the strong magnetic penetration ability, magnetic nanoparticles (MNPs) are widely utilized in magnetic measurement technologies in biomedicine[5]. However, the potential of magnetic nanoparticles (MNPs) in biomedical applications is limited by several challenges, including biotoxicity, stability issues, and the complexity involved in their preparation and functionalization[6]. Thus, replacing MNPs with other more superior magnetic materials presents a feasible alternative. In the field of biosensing, Lanthanide-binding tags (LBTs) are short peptides comprising of around 20 amino acids, exhibit unique physical properties due to their ability to bind various lanthanide elements (such as lanthanum, europium, terbium and gadolinium) in high-affinity[7]. Besides that, the small size offer LBT versatility to intergrade into other protein as the probe without changing the protein structure and function[8]. In biological detection, identify specific biomarkers with high sensitive and accuracy is a key challeng[9]. Lanthanide ions show strong magnetism due to their unpaired 4f electronic structure[10], which make them as the perfect candidates for quantum sensing and could overcome the background noise in fluorescent-based detection. By carrying out lanthanide ions, LBTs could utilize as a functional probe for biomarker detection. To enable the loading of LBT onto desired biomolecules, here we fused the LBT with specific truncated antibody, call nanobody, which dissociation constants with RBD is as low as 1.86 nM[11] Although LBT has demonstrated significant potential in biomolecular detection, its current application of primarily leverage the optical properties of lanthanide ions, while their magnetic properties remain largely unexplored.

Quantum sensings with the nitrogen-vacancy (NV) centers in diamonds, among the most thoroughly researched and well-understood spin defects, offer an ideal experimental platform[12]. An NV center consists of a nitrogen atom substituted in the carbon lattice next to a vacant site, as depicted in **Figure 1b**. The NV center has several valuable properties for quantum sensing. They have atomic size, which provides sub-nanometer resolution and allows them to be very close to sample[13]. Most importantly, their long spin lifetime at room temperature and exceptional biocompatibility make them ideal probes for biological applications[14]. As a quantum sensor, the NV centers can detect small magnetic fields and magnetic noise produced by nuclei spins or unpaired electrons[15].

In this paper, we propose a biomolecule detection scheme that exploits the relaxometer based on ensemble NV centers in diamonds and the magnetism of lanthanide ions. By engineering the diamond surface to specifically capture biomolecular targets and using LBTs as probes, a low-cost and background fluorescence-free immunomagnetic assay can be developed. First, we validated the capability of our setup to detect LBTs carried with lanthanide ions. Subsequently, we demonstrated the detection of the specific interaction between the SARS-CoV-2 spike protein receptor-binding domain (RBD) and its antibody. By varying the concentration of the input biomolecules, we observed a linear decrease in T1 values with increasing targets. We evaluated the detection limit is approximately 1 pmol. These results confirm the effectiveness of utilizing NV centers for biomolecule detection.

## 2. Results

### 2.1. Principle and Conceptual Design

**Figure 1a** depicts the conceptual design of our detection scheme. Initially, The diamond surface is silanized to introduce amine functional groups. Then, the diamond surface is coated with primary antibodies $NH_3^+$ to capture interested biomolecular targets. And the diamond surface would be passivation by bovine serum albumin (BSA) to occupy the unbind site to prevent non-specific binding. Next, the test samples are introduced, followed by a wash out step to remove the molecules do not bind to primary antibody. Subsequently, $Gd^{3+}$-containing LBTs are introduced. These engineered LBTs were fused with nanobody which can specifically recognized the targets. After another wash step, only the LBTs bound to the target will retain on the surface. The strong, fluctuating magnetic signals produced by unpaired electrons in $Gd^{3+}$ would then interact with the NV centers in the diamond substrate below. The interactions are monitored by observing changes in the T1 relaxation times of the NV centers. Therefore, the specific targets could be detected by the changing of T1 relaxation times.

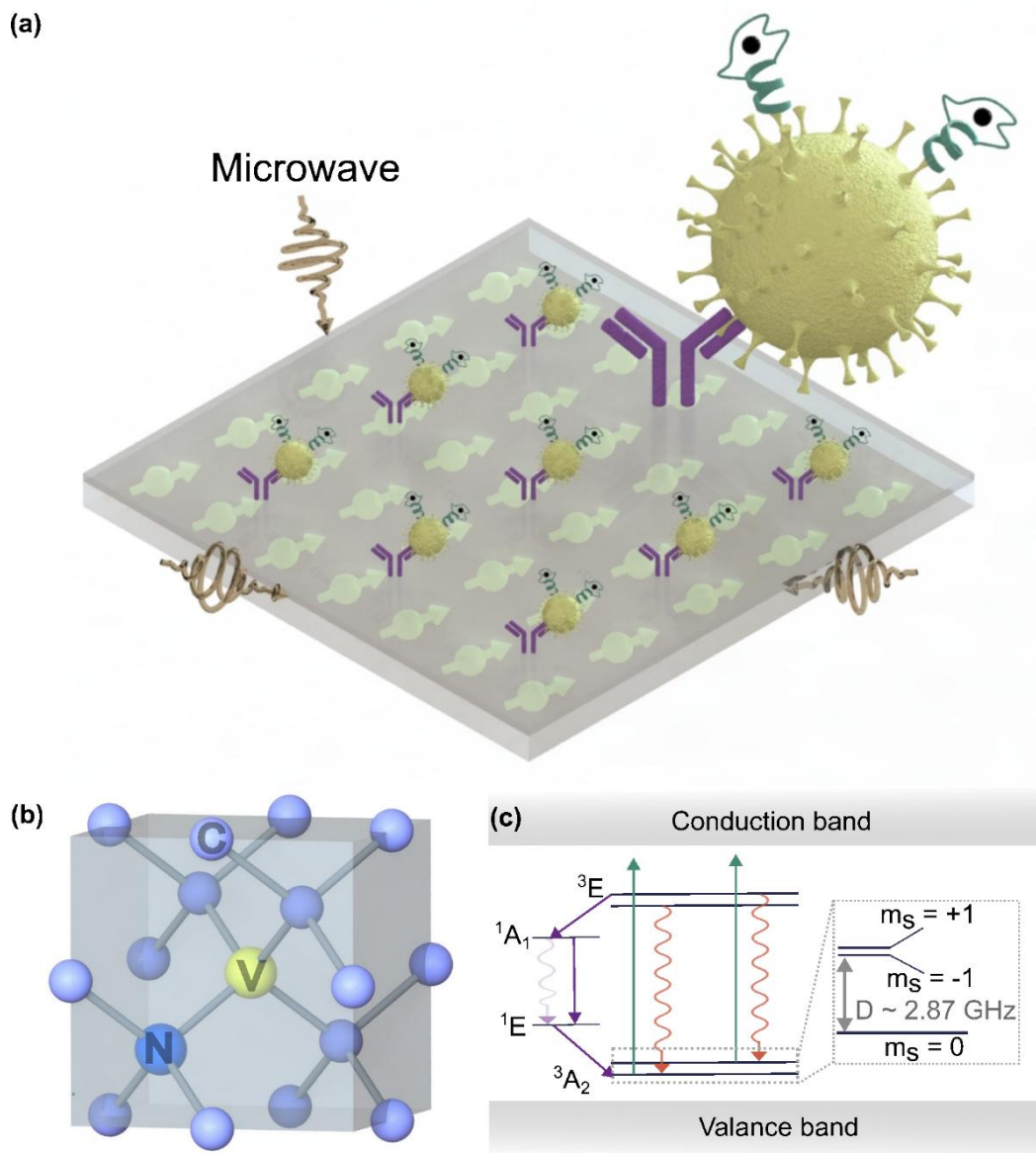

**Figure 1.** (Detection scheme and basic characteristics of nitrogen vacancy (NV) center. a, Proposed detection scheme: There are NV centers (light green) in the near surface of diamond (gray). Primary antibodies (purple) are engineered to attach to the diamond surface. The primary antibodies capture target biomolecules (yellow). LBTs fused with nanobody (blackish green) specifically bind to the target protein and retain on the surface. b, Atomic structure of an NV center. c, Simplified energy level diagram of NV center.)

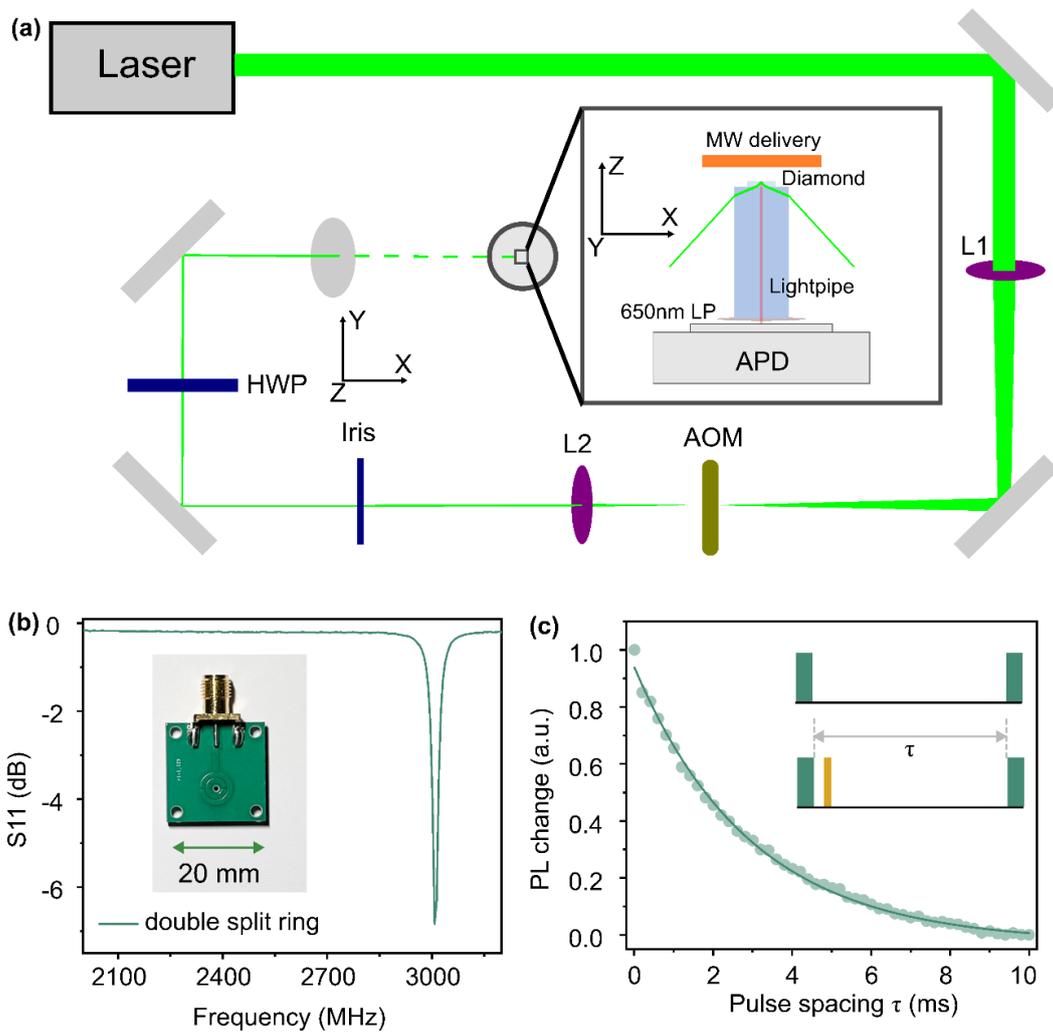

**Figure 2** (Experimental apparatus for the detection of quantum states in the NV center and related measurement results. a, Diagram of experiment setup. b, S11 of the antenna. c, Typical measurement results for relaxation time (T1) with the measurement sequence displayed in the inset. green blocks indicate laser pulses and the orange block represents a microwave pulse that shifts the population between the two states.)

The sensitivity of NV centers to magnetism stems from their unique energy structures. **Figure 1c** presents a simplified energy diagram of the NV center, which has a triplet ground state ($^3A_2$) and excited state ($^3E$). In the absence of a magnetic field, zero field splitting causes each state to divide into two energy levels, labeled $|m_s = 0\rangle$ and $|m_s = \pm 1\rangle$. The energy gap between these levels is approximately 2.87 GHz, which permits modulation of their populations using a resonant microwave pulse. Crucially, the NV centers can be initialized into the $|m_s = 0\rangle$ state using a green laser pulse lasting several microseconds. This optical initialization is facilitated by the spin-dependent intersystem crossing (ISC) rate of the NV centers. As indicated by the purple arrows in Figure 1c, upon excitation to the excited states, the $|m_s = \pm 1\rangle$ state is more likely to undergo intersystem crossing, subsequently decaying to the singlet state before returning to the $|m_s = 0\rangle$ ground state. In contrast, the $|m_s = 0\rangle$ state retains its quantum number $m_s$. Consequently, the NV center population is optically pumped into the $|m_s = 0\rangle$ state, achieving polarization levels up to 96%[16]. Moreover, the fluorescence intensity of the $|m_s = 0\rangle$ and $|m_s = \pm 1\rangle$ state is different due to the ISC rates. The $|m_s = \pm 1\rangle$ state is relatively darker than the $|m_s = 0\rangle$ because the emission of singlet state is filtered by our setup.

T1, employed for detecting LBTs, denotes the spin lifetime of the NV centers. Once polarized into the $|m_s = 0\rangle$ state by a green laser pulse, the NV centers gradually return to thermal equilibrium at room temperature, where the populations of $|m_s = 0\rangle$ and $|m_s = \pm 1\rangle$ are equal[17]. T1 measures the duration the NV centers remain in the polarized state. Details of the experiment setup of T1 relaxometry are shown in **Figure 2a**. Microwave was delivered with an antenna designed following the reference[18]. Measured S11 of the antenna and the antenna used to deliver microwave device are shown in **Figure 2b**. The measurement of T1 uses the pulse sequences depicted in the inset of **Figure 2c**. The sequences shown in the upper and lower parts of the figure are applied consecutively, with varying intervals between the initial and final laser pulses. The final signal is calculated using (SIG1 − SIG2)/(SIG1 + SIG2), where SIG1 and SIG2 are the fluorescence levels measured by the final laser pulses of the upper and lower sequences, respectively. This calculation helps to mitigate any fluctuations unrelated to changes in spin states. Figure 2c displays a typical T1 measurement, where increasing the time interval between pulses brings the difference between SIG1 and SIG2 closer to zero due to longitudinal relaxation. By fitting these results with a single exponential decay function, the T1 value can be extracted.

**2.2. Magnetic Detection of LBT Concentraion**

To confirm the capability of our setup in detecting LBTs, we engineered the diamond surface to capture the LBTs and measured T1. For detailed information on LBTs, please refer to supporting information. The process of surface treatment is shown in **Figure 3a** . Initially, the diamond is hydroxylated through a pickling process, followed by silanization using 3-Aminopropyltriethoxysilane (APTES, CAS 919-30-2) to introduce amino functional groups onto the surface. This treatment allows the diamond to adsorb biomolecules non-specifically. Specifically, to assess the diamond's response to LBTs, the silanized surface was exposed to solutions of LBTs at different concentrations.

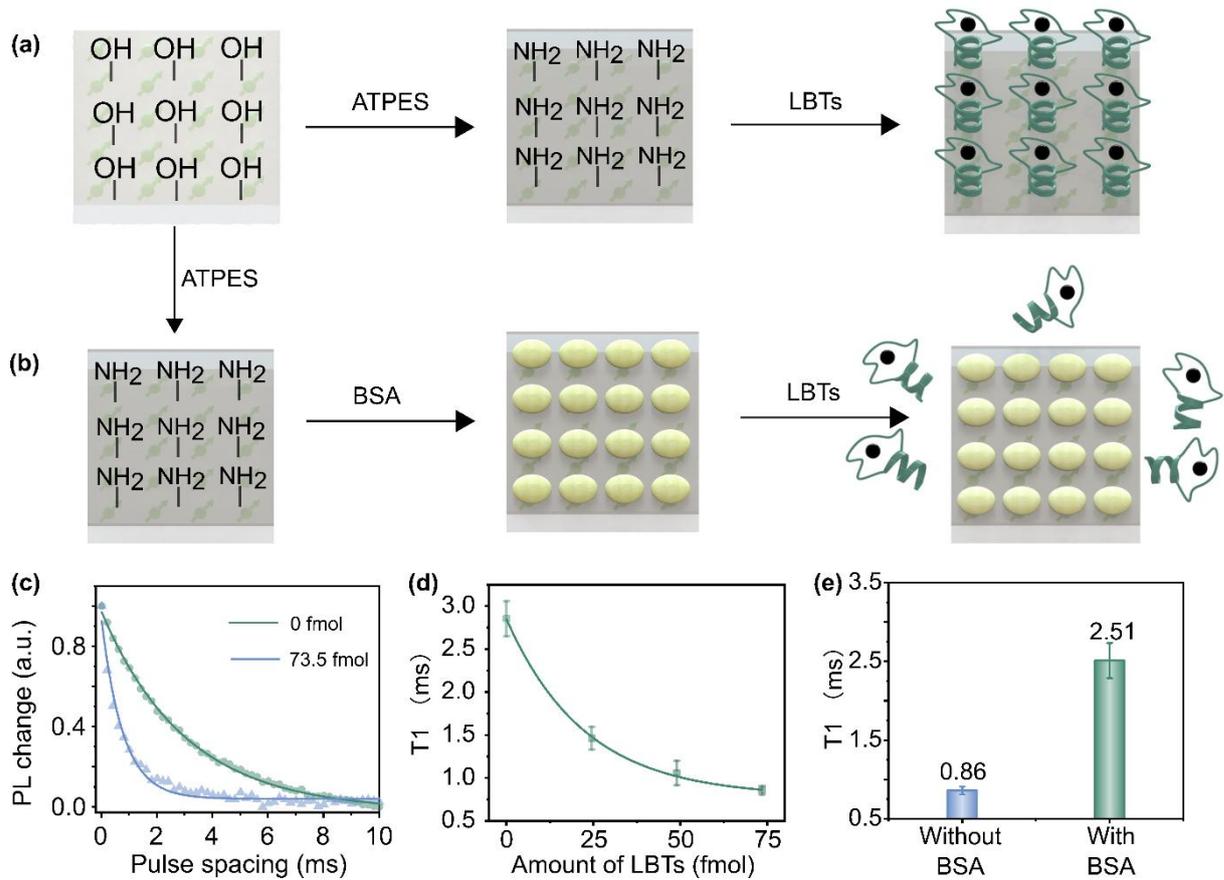

**Figure 3** (The process and results of NV central magnetic detection of LBTs. a, Process of surface engineering on diamond to detection of LBTs. b, BSA shielding of LBTs in the diamond surface engineering process. c, Comparison of T1 results for LBTs solutions at amounts of 0 fmol and 73.5 fmol. The data are fitted using a single exponential decay function, with T1 values fitted to 2.856 ms and 0.77 ms, respectively. d, T1 values corresponding to various amounts of LBTs. e, Assessment of the effectiveness of BSA blocking.)

With LBT attachment, the T1 relaxation time showed significant decrease, as shown in the **Figure 3c**. Without LBTs solution treatment, T1 is recorded at 3.02 ms. After the silanized diamond surface is fully exposed to the LBTs solution and thoroughly rinsed, T1 measurements of 0 fmol is taken as a reference. At LBTs amount of 73.5 fmol, T1 significantly drops to 0.77 ms, indicating a substantial adsorption of $Gd^{3+}$ on the diamond surface. This reduction in T1 suggests that the spin dipole field from the $Gd^{3+}$ ions within the LBTs disrupts the magnetic stability of the surface, causing fluctuations in the magnetic field detected by the nearby NV centers. The method uesd to calculate the amount of proteins is presented in SI.

Further T1 measurements with different LBTs amounts are presented in **Figure 3d**. To ensure data reliability, T1 values from five randomly selected locations on the diamond surface were averaged. These results demonstrate that the T1 relaxation time of the surface-near NV centers can detect changes in surface magnetic noise. As LBTs amount increases, T1 consistently decreases.

To minimize non-specific binding in subsequent detections, it is essential to passivate the diamond surface after it has been coated with antibodies. Following the functionalization of the diamond surface with desired biomolecules, the remaining unbound sites may unintentionally adsorb LBTs, resulting in the false positive. To prevent this, a commonly used surface passivation protein bovine serum albumin (BSA) was applied to block the empty binding sites of diamonds. The process is illustrated in **Figure 3b**.

To assess BSA's effectiveness, it was initially applied to the silanized diamond using a 3% solution. Following this, the diamond was exposed to LBTs solution, which would result in a detection of 73.5 fmol on the silanized surface if without BSA. Phosphate-Buffered Saline with 0.1% Tween 20 Detergent (PBST) buffer was applied after treatment of BSA and LBTs to wash out the unbind proteins. T1 results from BSA-treated surfaces maintained a value of approximately 2.51 ms, close to the baseline level with 0 fmol LBTs, compared to a drop to 0.86 ms without BSA blocking (**Figure 3e**). This indicates that BSA effectively prevents non-specific bindings.

**2.3. Immunomagnetic Assay Based on biomolecular interactions**

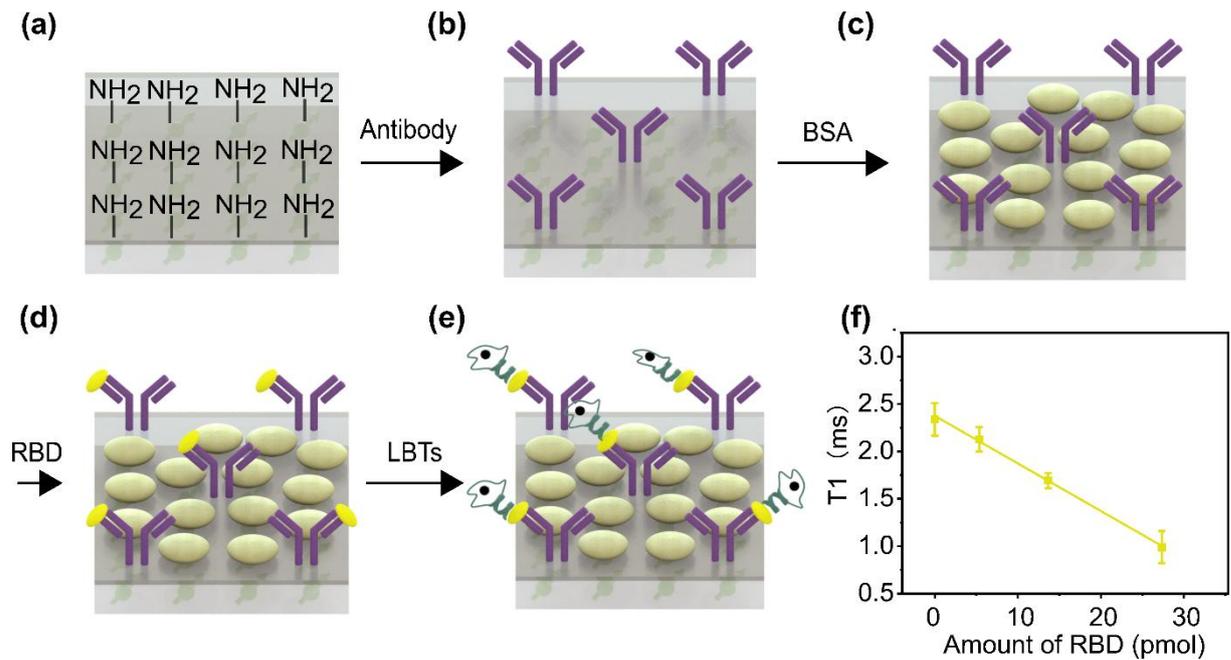

**Figure 4** (Process and results of LBTs Immunomagnetic assay based on NV centers. a - e, depict the sequential process of treating diamond surfaces. F, illustrates the T1 response to varying amounts of RBD applied in step (d).)

To verify the application of NV center as the biomarkers detector, we used the above scheme to detect the receptor-binding domain (RBD) of SARS-CoV-2 spike protein. As shown in **Figure 4a** to **4e**, the diamond was sequentially immersed to piranha solution and APTES to introduce $NH_2$ groups. Followed by coating the surface with RBD polyclonal antibodies via static electrostatic interactions. To prevent non-specific adsorption, BSA was applied to fill any remaining vacancies on the surface after antibody adsorption. The surface was thoroughly rinsed after each treatment step to remove any unbound proteins. Next, purified RBD was loaded onto the diamond and then washed out. The surface was treated with the LBTs solution. The fused anti-RBD nanobody allowed LBTs to target RBD specifically[11], forming an antibody-RBD-LBTs sandwich structure. After binding, the surface was rinsed to remove non-specifically bound RBD and LBTs. Finally, T1 was measured to assess the amount of specifically bound LBTs.

The response of the NV center's spin relaxation time is influenced by both the concentration of LBTs and their distance from the NV centers. However, our test does not need to account for the impact of this distance uncertainty on the veracity of the results. Since antibodies are immobilized on the diamond surface, and LBTs bind specifically to the antibodies via RBD. There's only one way LBTs can bind specifically to surfaces. So the distance variation between the specifically bound LBTs and the NV centers is kept within an acceptable range. Furthermore, we use an ensemble of NV centers, averaging out these limited distance variations in each measurement. Additionally, by conducting multiple time-series measurements and sampling across different regions, we further reduce the impact of distance uncertainty on the accuracy of determining biological target concentrations.

We evaluated the response of NV centers to different concentration of RBD, with results shown in **Figure 4f**. To ensure that the relaxometer responds exclusively to RBD, we employed oversaturated antibodies and LBTs. The results indicate a decrease in T1 with the increasing of RBD concentration. when no RBD is present (RBD = 0), after adding the LBT solution and subsequently washing the surface with buffer solution, the T1 relaxation time of the NV center remains nearly unchanged. This indicates that LBTs do not remain on the surface, even in the presence of other proteins such as primary antibodies, confirming the specific binding of LBTs to RBD. Furthermore, since the decrease in T1 relaxation time of the NV center is proportional to the number of LBTs, the observed linear relationship between T1 response and RBD concentration further supports the high selectivity of LBT binding to RBD.

**Table 1.** (List of different probe types, detection targets and detection limits)

| Probe | Type | Target | Limit of detection |
| --- | --- | --- | --- |
| LBTs (method used in this article) | Magnetic | RBD of SARS-CoV-2 spike protein | 1pmol |
| NaYF4:Yb,Er [19] | Optical | Arginine | 15.6 mM |
| NaGdF$_4$:Yb$^{3+}$, Er$^{3+}$@SiO$_2$–spiropyran [20] | Optical | Histidine | 4.4µM |

| Material | Type | Target | Detection limit |
|---|---|---|---|
| $Y(CH_3COO)_3 \cdot 4H_2O$ (X = Y, Yb, Er, Tm) [21] | Optical | Glutathione/$H_2O_2$ ratio | Glutathione 1mM $H_2O_2$ 5μM |
| $NaYF_4$:Yb,Tm@$TiO_2$ [22] | Optical | Carcinoembryonic antigen | 3.6 pg/mL |
| Type α-NaYF4:$Yb^{3+}$, $Er^{3+}$ [23] | Optical | $ClO^-$ | 16.1nM |
| Polyethyleneimine (PEI) as a mediate shell to combine superparamagnetic $Fe_3O_4$ core with dual quantum dot shells[24] | Optical | SARS-CoV-2 nucleocapsid protein | 0.012 ng/mL |
| Magnetic Nanoparticles (MNP)[25] | Magnetic | Mimic SARS-CoV-2 consisting of spike proteins and polystyrene beads | 0.084nM (5.9fmol) |
| Polyethyleneimine (PEI) as a mediate shell to combine superparamagnetic $Fe_3O_4$ core with dual quantum dot shells[24] | Magnetic | SARS-CoV-2 nucleocapsid protein | 0.235 ng/mL |
| Secondary antibody-conjugated MNP[5] | Magnetic | Anti-RBD antibody | ~1ng/mL |
| AuNPs-SPCE Aptasensor [26] | Electricity | Spike protein of SARS-CoV-2 | 1.30pM |
| PtNPs/GO-COOH Aptasensor[27] | Electricity | Alpha-fetoprotein | 1.22 ng/mL |
| Surface acoustic wave biosensor[28] | Acoustic | DNA | 14.4pM |
| Rayleigh wave - Shear horizontal surface acoustic wave[29] | Acoustic | Single-stranded DNA | 0.617ng/mL |

| Phase-Change microcapsule-carbon nanotube sensor[30] | Thermal | Glucose | 11.73-13.11µM |

The relationship between T1 and the RBD concentration fits a linear model with a slope of -0.05±0.0015 ms/pmol. Given an average measurement duration of one minute, the T1 measurement error is 0.1 ms, suggesting that our detection threshold for RBD is approximately 2 pmol. Table 1 demonstrates that the accuracy of magnetic detection using LBTs surpasses that of most optical detection methods employing lanthanide elements, and is comparale to the accuracy obtained with other magnetic particles. Compared to electrical, acoustic, and thermal sensing methods, this approach offers advantages such as low sample requirement, high signal-to-noise ratio (SNR), and high sensitivity. While its accuracy is not yet on par with mainstream virus detection methods such as RT-PCR, this technique offers notable benefits including rapid testing, high detection efficiency, low cost, and reusability.

It is hypothesized that the diminished sensitivity observed in comparison to LBTs detection may be due to the increased distance between the LBTs and the NV center, caused by the sandwich structure of the antibody and RBD, which lengthens the distance and leads to a weakened response signal. With the simple set up, this method can be easily adapted and modified to detect multiple target proteins by replacing the functionalization antibody on the diamond surface and the nanobody fused with LBTs. Enhancements in sensitivity could be achieved by strategically redesigning the antibody-RBD-LBTs sandwich structure to be shorter, which would enhance the NV center response and increase the T1-RBD slope. Alternatively, using nanodiamonds, which have a larger surface-to-volume ratio, could also improve sensitivity.

## 3. Conclusion

In summary, we proposed a method to detect protein targets such as SARS-CoV-2 spike protein by employing a relaxometer based on diamond NV centers. This method offers the advantages of being low cost and free from background noise, utilizing the magnetic properties of LBTs in conjunction with the quantum characteristics of NV centers. As a demonstration, we performed experiments measuring the T1 of NV centers in response to varying amounts of the RBD of the SARS-CoV-2 spike protein, establishing a detection limit of approximately 1 pmol. This technique is not only applicable to the detection of SARS-CoV-2 spike protein as demonstrated, but also readily adaptable for the identification of other biomolecule detections. Our method is significant in advancing the field of magnetic biosensing, providing a robust platform for the sensitive detection of diverse biological entities in medical diagnostics and research.

## 4. Experimental Section/Methods

*Purification of LBT*: The LBT[8] and RBD-nanobody[11] fusion protein was synthesized by BioBasic into pET-28a(+) vector, with 6×His and MBP tag at N-terminal. The expression vector was transformed into *Escherichia coli* BL21 (DE3) Rosetta strain and cultured at LB plate with kanamycin resistance overnight. Single colony was picked up and cultured in Terrific Borth. Isopropyl β-d-1-thiogalactopyranoside (Thermo Scientific) was added when bacteria culture reach OD600 1.0 and continue culture for overnight at 20℃, 180 rpm. Bacteria was harvested and resuspended by binding buffer containing 20 mM HEPES (20 mM, CAS 7365-45-9), 500 mM NaCl (CAS 7647-14-5), 20 mM imidazole (CAS 288-32-4), and a pH of 7.4. One tablet of Pierce Protease Inhibitor (Thermo Scientific) and 1 mM phenylmethylsulfonyl fluoride (CAS 329-98-6) was added to prevent protein degradation. Bacteria cells were lysis by LM20 microfluidizer. After centrifugation to remove the pellet and filtered by 0.45 μm membrane, the supernatant of cell lysis was loaded to a 5 mL HisTrap column (GE Healthcare) and controlled by an ÄKTA™ FPLC system (GE Healthcare). Protein was eluted by elution buffer (20 mM HEPES, 500 mM NaCl, 500 mM imidazole, pH 7.4). Eluted protein was characterized by sodium dodecyl sulfate–polyacrylamide gel electrophoresis (SDS-PAGE). Protein elution which contained the target protein band was collected and further loaded onto the HiLoad 16/600 Superdex 200 pg column (GE Healthcare) with gel filtration buffer, which contains HEPES (20 mM), NaCl (500 mM), and 10% glycerol (CAS 56-81-5) at a pH of 7.4. Target protein band was characterized by SDS-PAGE and collected for further experiment.

*Operational procedures for LBTs combined with $Gd^{3+}$*: LBT protein was mixed with 2 mM $GdCl_3$ (CAS 10138-52-0) buffer and vortexed for 1 min. The LBT protein would precipitate after binding with Gd ions. The LBT protein with $Gd^{3+}$ mixture was centrifuged to remove the un-binding $Gd^{3+}$, with using PBS buffer to resuspend and wash the pellet with 3 times repeat.

*Glass container cleaning process* : Before conducting experiments using glass containers, it is crucial to follow a thorough cleaning protocol to maintain cleanliness. Begin by placing the container in an ultrasonic cleaning machine and clean it with acetone (CAS 67-64-1) for 5 minutes, followed by drying with a blast of air. Subsequently, clean the container again in the ultrasonic cleaning machine using isopropyl alcohol (IPA, CAS 67-63-0) for another 5 minutes and dry it with air. It is important to use the containers promptly after cleaning to prevent any recontamination.

*Diamond surface cleaning process*: Place the diamond in 2M NaOH (CAS 1310-73-2) solution for ultrasonic cleaning for 5-10 minutes. After cleaning, Transfer the diamond to a pickling solution composed of sulfuric acid (CAS 7664-93-9) and nitric acid (CAS 7697-37-2) in a 1:1 ration, adding the sulfuric acid before the nitric acid, maintain the mixture at 210 °C for 8 hours. After pickling, carefully dispose of the pickling solution and thoroughly rinse 3 times with deionized water. Finally, clean with acetone in an ultrasonic cleaner for 5 minutes, air dry, then clean with IPA for another 5 minutes, and air dry again.

*Diamond surface hydroxylation process*: Place the cleaned diamond into the piranha solution, prepared by mixing sulfuric acid and hydrogen peroxide (CAS 7722-84-1) in a 3:1 ratio, adding sulfuric acid first followed by hydrogen peroxide, and seal the container. Then heat it at 75°C for 45 minutes. After heating, carefully dispose of the piranha solution and wash the diamond three times with an ethanol (CAS 64-17-5) solution. Finally, air-dry the diamond and further dry it in an oven at 60°C.

*Diamond surface silanization process*: Place the hydroxylated diamond into the ATPES solution (2% APTES (CAS 919-30-2) and 98% toluene (CAS 108-88-3)), then heat it in a sealed environment at 50°C for 16 hours. After heating, carefully dispose of the APTES solution and wash the diamond 3 times with ethanol. Next, subject the diamond to ultrasonic treatment in ethanol for 10 minutes to remove any silane residues not covalently bonded to the surface. After ultrasonic treatment, wash the diamond again with ethanol. Finally, dry the diamond at 60°C.

*Sample treatment of magnetic detection of different concentrations of LBTs*: First, prepare the gel filtration buffer solution (buffer component: HEPES (20mM), NaCl (500mM), 10% Glycerol (v/v %). Used NaOH and HCl to adjust pH to 7.4, and used 0.45 μm filter to filter the buffer. Next, dilute the LBTs solution to the required concentrations using gel filtration buffer). The silanized diamond is then placed in these varying concentrations of LBTs solutions and incubated at room temperature for 1 hour with agitation. After incubation, remove the diamond from the solution and wash it three times with PBST buffer (1×PBS buffer containing 0.1% Tween 20 at pH7.4 (CAS 9005-64-5)) to eliminate any un-adsorbed LBTs. Finally, the diamond is optically tested.

*Sample treatment of immunomagnetic assay based on biomolecular interactions*: First, dilute the antibody solution to the desired concentration using 1×PBS buffer (pH 7.4). Place the silanized diamond in the prepared antibody solution and incubate at room temperature with shaking for 1 hour. After incubation, wash the diamond three times with PBST buffer. Next, dissolve BSA protein in 1×PBS buffer (pH 7.4) to the required concentration and incubate the diamond in this BSA solution under the same conditions.

Following the BSA treatment, wash the diamonds again three times with PBST buffer. Then, dilute the RBD solution using 1xPBS buffer (pH 7.4) to different concentrations, incubate the diamond in the RBD solution for 1 hour at room temperature, and then wash three times with PBST buffer. Prepare the LBTs solution at required concentration using gel filtration buffer, Incubate the diamonds in this solution for 1 hour with shaking at room temperature, and wash three times with PBST buffer. Finally, the diamond is optically tested.

*Optical Components*: To enhance the collection efficiency of NV fluorescence and facilitate total internal reflection, we utilized a hexagonal lightpipe (#49-402, 4mm Aperture, 50mm Length Standard NA, Edmund Optics) to channel the fluorescence from the diamond to an amplified photodiode (A-CUBE-S3000-03, Laser Components). The experimental setup includes a green laser beam (Sprout-H 10W) that is redirected by a mirror (PF10-03-P01, Thorlabs) and focused through a convex lens (L1 and L3, LA1074-ML, Thorlabs) with a focal length of 20 mm. The iris (D20S) adjusts the intensity of the laser beam, while the Acousto-Optic Modulator (AOM, G&H AOMO 3200-121) controls the beam's on/off switching. The Half-Wave Plate (HWP,) is used to control and manipulate the polarization of laser beams. The beam is refracted by the lightpipe's side, travels along the lightpipe, and focuses on the diamond's upper surface. Total internal reflection occurs at this surface, preventing the optical bleaching of biomolecules on the diamond. The reflected beam exits the lightpipe and the fluorescence, after undergoing total internal reflection within the lightpipe, is directed into the amplified photodiode following filtration through a 650 nm long pass filter (FELH0650, Thorlabs). The photodiode then converts the fluorescence intensity into analog voltage signals.

*Data acquisition*: A data acquisition system (DAQ, USB 6343 BNC, National Instruments) was utilized to capture the analog signal from the output of the avalanche photodiode and convert it into digital data.

**Supporting Information**

Supporting Information is available from the Wiley Online Library or from the author.

**Author Contributions**

Zibo Gao, Zhengzhi Jiang, and Qiyu Liang conceptualized the study, contributed to methodology, handled materials, and participated in all writing tasks. Weibo Gao, Yansong Miao, Ruihua He, Hongliang Sun and Mo Li, supervised the project. Van Cuong Mai, Yingwei Tang, Qirong Xiong, Wenting Zhao and Hongwei Duan assisted in developing the methodology. Zibo GAO, Zhengzhi Jiang and Qiyu Liang contributed equally to this work

**Acknowledgements**

This work is supported by Singapore Quantum engineering program (NRF2021-QEP2-03-P01, NRF2021-QEP2-03-P10, NRF2021-QEP2-03-P11), ASTAR (M21K2c0116, M24M8b0004), Singapore National Research foundation (NRF-CRP22-2019-0004, NRF-CRP30-2023-0003, NRF2023-ITC004-001 and NRF-MSG-2023-0002) and Singapore Ministry of Education Tier 2 Grant (MOE-T2EP50221-0005, MOE-T2EP50222-0018).

**Conflict of Interest Statement**

The authors declare no conflict of interest.

**Data Availability Statement**

The data that support the findings of this study are available from the corresponding author upon reasonable request.

**Ethical Statement**

Received: ((will be filled in by the editorial staff))

Revised: ((will be filled in by the editorial staff))